\documentclass[aps,amsfonts,amsmath,prd,preprint,nofootinbib]{revtex4}
\newcommand{\beq}{\begin{equation}}
\newcommand{\eeq}{\end{equation}}

\usepackage{epsfig,bbm,cancel,ulem}
\usepackage[breaklinks=true]{hyperref}

\newcommand{\handhika}[1]{\textit{#1}}

\newcommand{\mc}[1]{\mathcal{#1}}

\begin{document}

\title{On Bogomol'nyi Equations of Classical Solutions}

\author{Ardian N. Atmaja}
\email{ardian\_n\_a@um.edu.my}
\affiliation{Quantum Science Centre, Department of Physics, Faculty of Science, University of Malaya, 50603 Kuala Lumpur, Malaysia.\\Research Center for Physics, Indonesian Institute of Sciences (LIPI), Kompleks PUSPIPTEK Serpong, Tangerang 15310, Indonesia.}
\author{Handhika S. Ramadhan}
\email{hramad@ui.ac.id}
\affiliation{Departemen Fisika, FMIPA, Universitas Indonesia, Depok, 16424, Indonesia. }
\def\changenote#1{\footnote{\bf #1}}

\begin{abstract}
We review the Bogomol'nyi equations and investigate an alternative route in obtaining it. It can be shown that the known BPS equations can be derived directly from the corresponding Euler-Lagrange equations via separation of variables, without having to appeal to the Hamiltonian. We apply this technique to the Dirac-Born-Infeld solitons and obtained the corresponding equations and the potentials. This method is suitable for obtaining the first-order equations and determining the allowed potentials for noncanonical defects. 
\end{abstract}

\maketitle
\thispagestyle{empty}
\setcounter{page}{1}

\section{Introduction}

Bogomol'nyi equation is a powerful tools in the study of solitons. It reduces the second-order equations into first-order. Such reduction greatly simplifies the problem, although in some cases the analytic solutions are still yet known. In his classic paper~\cite{bogol}, Bogomol'nyi showed that the field equations for topological defects (see, for instance~\cite{vilenkinshellard}) can be reduced to first-order provided the coupling constant of its potential takes certain values namely Bogomol'nyi-Prasad-Somerfield (BPS) limit, domain walls being the only exception where all solitonic solutions are also Bogomol'nyi solutions. The solutions can be identified as approaching the global minima at the boundary, resulting in the saturation of the corresponding static energy, where each point at the boundary maps to different global minimum and thus guaranteed to be stable. In the supersymmetric extension of the theories, the Bogomol'nyi solutions correspond to ``short'' supermultiplets which are invariant under some parts of supersymmetry transformation and they are usually called Bogomol'nyi-Prasad-Sommerfield (BPS) states~\cite{weinberg}. 

Recently, there have been many models of topological defects with noncanonical terms~\cite{babichev1, sarangi, babichev2, pavlovsky, ramadhan}. One of the models extensively studied are the Dirac-Born-Infeld (DBI) type models, inspired by the non-linear electromagnetic~\cite{borninfeld, boillat}\footnote{Once proposed to regularize the divergence of the electron's self-energy.} which for the last three decades regains interest due to the progress of $D$-branes in string theory~\cite{callan, gibbons, arkady}. The nonlinearity of the DBI kinetic terms can evade the stringent constraint by Derrick on the existence of finite-energy solutions~\cite{derrick}. In the context of cosmology such defects might have been formed during the phase transitions at the stringy scale. Most of the proposals focus on the exact solutions obtained from the full second-order differential equations. In particular, in~\cite{babichev2} all solutions of the Dirac-Born-Infeld (DBI) gauge vortices have energies greater than the BPS state, which so far cannot be obtained~\cite{babichevprivate}. Finding the first-order equation is thus essential, not only because it is interesting in its own right, but also because the solutions that saturate such equation can be identified as noncanonical solitons with the least energy. These solitons might have different properties than its canonical BPS counterparts.  

To the best of our knowledge, such attempt was first proposed in~\cite{Shiraishi}\footnote{See also~\cite{yisong}.}, where the authors discussed DBI vortices and beautifully showed that the first-order equations exist provided the potential is not an arbitrary symmetry-breaking shape but instead takes some specific form. This limits the availability of BPS solutions in DBI theories. It is surprising that the nonlinearity of kinetic term imposes such stringent constraint on the shape of the potential allowed. Later in~\cite{hadronic} one of the author applied the same technique to obtain BPS Born-Infeld monopoles and instantons, where both the xYang-Mills and Higgs fields are under the square-root. 

It is therefore intriguing to find a general mechanism in obtaining Bogomol'nyi equations. It is our goal in this paper is to report a preliminary study on such mechanism. In the following section, we review the Bogomol'nyi formalism from the point of view of global minimum of the Hamiltonian. Next, we will construct our formalism, a method of achieving first-order equations directly from the Euler-Lagrange equations, which we dubbed the {\it on-shell} method. It is the main result of our work. The next sections are devoted to applying it to obtain Bogomol'nyi equations for canonical defects. As a warm-up, we start with domain walls then proceed to Abelian vortices~\cite{abri} and non-Abelian magnetic monopoles~\cite{prasad}, where we show that they can be obtained via separation of variables of the auxiliary constraints. We will also show that the Bogomol'nyi equations for DBI domain walls~\cite{yisong} and vortices~\cite{Shiraishi} are consequences of the same condition. Finally we will discuss the possibility of obtaining BPS solutions in noncanonical defects in general.

\section{Off-shell method: Squaring The Energy Density}

Let us consider a Lagrangian density of $N_\phi$ fields\footnote{It should be noted that we do not restrict the fields to be only scalar. $\phi^{a}$ may represent scalar as well as vector fields, as we shall see later.} as follows
\begin{equation}
 \mc{L}\equiv \mc{L}(\phi^a, \partial_\mu\phi^a),
\end{equation}
where $a=1,\ldots,N_\phi$ and $\mu=0,\ldots,D$ is the Minkowskian spacetime index in $D+1$ dimensions with mostly positive signature, $(-,+,+,\ldots)$. The equations of motion are given by Euler-Lagrange equations
\begin{equation}
 {\partial{\mc{L}} \over {\partial\phi^a}}-\partial_\mu{\partial{\mc{L}}\over\partial(\partial_\mu\phi^a)}=0.
\end{equation}
Hamiltonian density can be derived from this Lagrangian density,
\begin{equation}
 \mc{H}=\sum_a\partial_t\phi^a \pi^a_\phi-\mc{L},\ \ \ \ \ \pi^a_\phi\equiv{\partial \mc{L} \over \partial(\partial_t\phi^a)}.
\end{equation}
For static fields, $\partial_t\phi^a=0$, it implies $\mc{H}=-\mc{L}$. The usual method in finding the Bogomol'nyi equations utilizes the Hamiltonian density for static solutions. It requires to rewrite the Hamiltonian density in the positive definite terms so that
\begin{equation}
\label{eq:hamiltonian}
 \mc{H}=\mc{A}^2_1(\phi, \partial_i \phi)+\mc{A}^2_2(\phi, \partial_i \phi)+\ldots,
\end{equation}
with $i=1,\ldots,D$ is the space index and $\phi\equiv(\phi^1,\ldots,\phi^{N_\phi})$. It then follows that the Bogomol'nyi equations are given by
\begin{equation}
\label{eq:positivedefinite}
 \mc{A}_1(\phi, \partial_i \phi)=0,\ \ \ \ \ \mc{A}_2(\phi, \partial_i \phi)=0,\ \ldots.
\end{equation}
The solutions are nonetheless the extremum points of the Hamiltonian (or Lagrangian) providing a condition that the remaining terms in the Hamiltonian, which cannot be written as positive definite terms, must contribute to the variation of the Hamiltonian (or Lagrangian) as boundary terms upon substituting the Bogomol'nyi equations. 

This method, however, is not always easy to execute because one has to make sure that the number of positive definite terms must be equal to the number of fields. Moreover, completing the squares can sometimes be cumbersome when the terms in the Lagrangian take noncanonical forms. Overcoming these difficulties is precisely our aim in looking for an alternative route.

\section{On-shell method}
Dealing with difficulties of the previous off-shell method, we propose a more rigorous way in finding the Bogomol'nyi equations for static solutions which takes a  different approach. Here we are going use what we call an {\it on-shell} method, by exploiting the Euler-Lagrange equations. An advantage of this method is that one is already working in the extremum points of the Hamiltonian (or Lagrangian) without worrying about the boundary terms. Another advantage is that the number of equations is equal to the number of fields by default. The method consists of few steps described below:\\
\begin{enumerate}
  \item \underline{One dimensional effective Euler-Lagrange equations}\\
We limit our case where the Euler-Lagrange equations can be written effectively as one dimensional second-order differential equations. For example one can consider the fields to be spherically symmetric in $D$-spatial dimension where $\phi^a\equiv\phi^a(r)$. For each field $\phi^a$ we write a general form of one dimensional effective Euler-Lagrange equation as follows
\begin{align}
 \partial_r\left[f^a(r,\phi)\partial_r\phi^a(r)\right]=g^a(r,\phi),
\end{align}
where $r$ is the one dimensional parameter. Note that repeating index in the above equation does not mean a summation and we implicitly write all the couplings in the equation above. In general, having first derivative fields dependent on the function $f$ could gives us indefinite form of equations as we will discuss in detail in the next step. However in some cases it is possible to have $f\equiv f(\partial_r\phi)$ and still obtain a definite form of equations such as some examples of Dirac-Born-Infeld action discussed in the next sections.
 \item \underline{Inserting auxiliary functions}\\
We modify the effective Euler-Lagrange equations by inserting some auxiliary functions that depend on the fields as such
\begin{equation}
 \partial_r\left[f^a(r)\partial_r\phi^a(r)-X^a(\phi)\right]=g^a(r,\phi)-X'^a(\phi),
\end{equation}
with $'\equiv{\partial\over\partial_r}$. Our prescription requires that the auxiliary functions must not contain any derivative of the fields, for otherwise we will not be able to achieve our goal in deriving the Bogomol'nyi equations. In principle we could have $X^a\equiv X^a(r,\phi)$ but in this article we take a simple case where no explicit dependence over the parameter $r$ in order for solving the auxiliary functions explained in the next steps. As we have learned, many examples of the well-known Bogomol'nyi equations posses a similar form.
 \item \underline{Identifying Bogomol'nyi and constraint equations}\\
From the previous step, we identify the Bogomol'nyi equations to be
\begin{equation}
 f^a(r)\partial_r\phi^a(r)=X^a(\phi),
\end{equation}
which are the solutions to the effective Euler-Lagrange equations providing that the constraint equations,
\begin{equation}
X'^a(\phi)= g^a(r,\phi),
\end{equation}
are satisfied. Since our main objective here is to find subset solutions of the second-order differential equations (the effective Euler-Lagrange equations) in terms of first-order differential equations (the Bogomol'nyi equations) we therefore require the constraint equations to be in the first-order equations as well. This requirement can not be fulfilled if $X^a\equiv X^a(\partial^n_r\phi)$, with $n>0$.
 \item \underline{Solving the auxiliary functions}\\
The next step is to find what auxiliary functions that solve both the Bogomol'nyi and constraint equations. In order to do that we write the constraint equations as follows
\begin{eqnarray}
\label{eq:algebraic}
 \sum_b {\partial X^a\over \partial\phi^b}\phi'^b&=& \sum_b {\partial X^a\over \partial\phi^b}{X^b\over f^b}\nonumber\\
						  &=&g^a(r,\phi),
\end{eqnarray}
where the right hand side of the first line is obtained by substituting the Bogomol'nyi equations. In this way we have transformed the constraint equations into algebraic equations\footnote{What we really mean is that the constraint equations do not contain fields' derivatives over $r$.} in terms of parameter $r$. This algebraic equations would be very complicated if $f\equiv f(\phi')$. For example in a theory with only one field we could have indefinite or iterated function  
\begin{equation}
\phi' ={X \over f\left(r,\phi,\phi'\right)}={X \over f\left(r,\phi,{X \over f\left(r,\phi,\phi'\right)}\right)}={X \over f\left(r,\phi,{X \over f\left(r,\phi,{X \over f\left(r,\phi,\ldots\right)}\right)}\right)}
\end{equation}
in order to write the right hand side of equation to be independent to the first derivative of the field for our method to work. However in principal we should be able to simplify it. For the above example, we could take an ansatz that $\phi'=Y(r,\phi)$ and so the equation becomes
\begin{equation}
 Y f(r,\phi,Y)=X,
\end{equation}
an algebraic equation. If it can be solved such that $Y\equiv Y(r,\phi,X(\phi))$ we may then proceed to the next step. For many fields, the equations we need to solve are
\begin{equation}
  Y^a f^a(r,\phi,Y)=X^a,\ \ \ \ \ a=1,\ldots,N_\phi.
\end{equation}
The main key in our on-shell method is that we make a conjecture that the consistent non-trivial solutions to the algebraic equations (\ref{eq:algebraic}) could be obtained by solving them order by order in the power of parameter $r$. This means that the algebraic equations (\ref{eq:algebraic}) must be in the form of explicit polynomial function of parameter $r$. Another crucial assumption we impose is that the auxiliary functions to be separable, that is $X=\prod_a X_a(\phi^a)$. As we shall see, all known Bogomol'nyi equations possess this behavior. 

\item \underline{Determining the topological charge}\\
As was pointed out by Wereszczynski~\cite{Wereszczynski}, the apparent ``drawback" in this on-shell method is perhaps that it does not directly gives the topological charge. However, such concern should not worry us. The charge can be obtained simply by inserting the BPS solutions into the functional static energy. 

Since
\begin{equation}
\label{eq:thisisenergy}
E=\int d^{D}x{\cal H}
\end{equation}
and since the correct BPS auxiliary functions $X^{a}(\phi)$ leads to (\ref{eq:positivedefinite}), then what is leftover in (\ref{eq:hamiltonian}) can be written proportionally as a total derivative of ``something"
\footnote{This is reminiscent of the self-duality method proposed by Adam {\it et al}~\cite{adam}. We thank Andrzej Wereszczynski for bringing that paper into our attention.}, 
\begin{equation}
{\cal H}={\cal H}_{BPS}\equiv\frac{1}{\sqrt{g_{rr}}}Q'.
\end{equation}
The BPS energy then gives
\begin{eqnarray}
 E_{BPS}&=&\int d^{D}x\ {\cal H}_{BPS}\nonumber\\
 &=&\int dr \sum^{N_\phi}_{a=1} Z_a(\phi) \phi'^a,
\end{eqnarray}
where we define $dQ(\phi)\equiv\sum^{N_\phi}_{a=1} Z_a(\phi) d\phi^a$. This ``something" $Q$ then depends only on the topology of the fields at the boundary which, if exists, it can be associated with the topological charge
\begin{eqnarray}
 E_{BPS}&=&Q\bigg |^{\phi^{a}_{max}}_{\phi^{a}_{min}}\nonumber\\
 &=&Q(\phi(\infty))-Q(\phi(0)).
\end{eqnarray}

We will show in the next sections how our on-shell method and the conjecture work to reproduce some well-known Bogomol'nyi equations in more detail. 

\end{enumerate}

\section{BPS Domain Walls}

Having laid down the mathematical formalism, we are ready to test it to the known cases. The simplest topological defects is given by the following Lagrangian
\begin{equation}
{\cal L}=\frac{1}{2}\partial_{\mu}\phi\partial^{\mu}\phi-V(\phi),
\end{equation}
with $\phi(x)$ a real scalar field and $V(\phi)=\frac{\lambda}{4}\left(\phi^{2}-\eta^{2}\right)^{2}$. The field equation is
\begin{equation}
\label{eq:domainwalls}
\phi''=\frac{\partial V}{\partial\phi}.
\end{equation}

It is known that the static energy can be written as~\cite{bogol}
\begin{eqnarray}
\label{eq:bogoleq}
E&=&\frac{1}{2}\int dx \left(\phi'\mp\sqrt{2 V}\right)^{2}\pm\int \sqrt{2 V}d\phi,\nonumber\\
&\geq& \left|\int \sqrt{2 V}d\phi \right|.
\end{eqnarray}
The bound is saturated by solutions that satisfy
\begin{equation}
\phi'=\pm\sqrt{2 V}.
\end{equation}
This is the Bogomol'nyi equation for domain walls. 

Let us now apply our method. Eq.(\ref{eq:domainwalls}) can be recast as
\begin{equation}
\label{eq:ourmethod}
\left(\phi'-X\right)'+X'=\frac{\partial V}{\partial\phi},
\end{equation}
with $X(\phi)$ an auxiliary constraint. It can then be written as a first-order equation,
\begin{equation}
\phi'=X,
\end{equation}
provided $X$ satisfies
\begin{equation}
X'=\frac{\partial V}{\partial\phi}.
\end{equation}
It is trivial to show that the solution for X is
\begin{equation}
X=\sqrt{2V}+c,
\end{equation}
for some constant $c$. Thus, we recover Eq.(\ref{eq:bogoleq}) if we set $c=0$.

\section{Bogomol'nyi Equations in Gauge Defects}

We can apply our formalism in Eq.(\ref{eq:ourmethod}) to defects possessing continuous symmetry, {\it e.g}, Abelian vortices and magnetic monopoles.

\subsection{Vortices}

Vortex is a two-codimensional defect that forms when the vacuum manifold is non-simply connected. It can appear in the Abelian-Higgs model with complex scalar field whose Lagrangian is given by, for example in~\cite{vilenkinshellard},
\begin{equation}
{\cal L}=\left(D_{\mu}\phi\right)^{*}\left(D^{\mu}\phi\right)-\frac{1}{4}F_{\mu\nu}F^{\mu\nu}-V(|\phi|).
\end{equation}
Employing the cylindrically-symmetric ansatz~\cite{abri},
\begin{eqnarray}
\phi({\bf x})&=&e^{in\theta}f(r),\nonumber\\
A_{\theta}({\bf x})&=&-\frac{n\alpha(r)}{er^{2}},\ \ A_{0}=A_{r}=0,
\end{eqnarray}
leads to the following static energy
\begin{equation}
E=2\pi\int dr\ r\left[f'^{2}+\frac{n^2\alpha'^{2}}{2e^2r^{2}}+\frac{n^{2}f^{2}\left(\alpha-1\right)^{2}}{r^{2}}+V\right],
\end{equation}
where we set the potential to be general. We wish to show later that the specific function of the potential emerge as a solution in our formalism. Rescaling $r\rightarrow er$ yields
\begin{equation}
\label{eq:Evortex}
E=2\pi\int dr\ r\left[f'^{2}+\frac{n^2\alpha'^{2}}{2r^{2}}+\frac{n^{2}f^{2}\left(\alpha-1\right)^{2}}{r^{2}}+{1\over e^2}V\right],
\end{equation}

Varying the static energy, we obtain the field equations
\begin{eqnarray}
\frac{(r f')'}{r}-\frac{n^{2}\left(\alpha-1\right)^{2}f}{r^{2}}-\frac{1}{2e^{2}}\frac{\partial V}{\partial f}&=&0,\nonumber\\
r\left(\frac{\alpha'}{r}\right)'-2f^{2}\left(\alpha-1\right)&=&0.
\end{eqnarray}
We can cast them into
\begin{eqnarray}
\frac{(rf'-X)'}{r}+\frac{X'}{r}&=&\frac{n^{2}f\left(\alpha-1\right)^{2}}{r^{2}}+\frac{1}{2e^{2}}\frac{\partial V}{\partial f},\nonumber\\
r\left(\frac{\alpha'}{r}-Y\right)'+rY'&=&2f^{2}\left(\alpha-1\right),
\end{eqnarray}
where $X$ and $Y$ are two auxiliary constraints, and in general $X=X(f, \alpha, r)$, $Y=Y(f, \alpha, r)$. We can collect our first-order equations
\begin{eqnarray}
\label{eq:bogolvortices}
rf'&=&X,\nonumber\\
\frac{\alpha'}{r}&=&Y,
\end{eqnarray}
which are guaranteed to be satisfied, provided 
\begin{eqnarray}
\label{eq:constraintsvortice1}
\frac{X'}{r}&=&\frac{n^{2}f\left(\alpha-1\right)^{2}}{r^{2}}+\frac{1}{2e^{2}}\frac{\partial V}{\partial f},\nonumber\\
Y'&=&\frac{2f^{2}\left(\alpha-1\right)}{r}.
\end{eqnarray}
Eqs.(\ref{eq:bogolvortices}) constitute our Bogomol'nyi equations, while conditions (\ref{eq:constraintsvortice1}) are the corresponding equations of constraints. 

In general, the functional $X$ and $Y$ can take any function of $f$, $\alpha$, and $r$. Here let us consider them to be independent of $r$ and take ansatz of separation of variables, {\it i.e.}, 
\begin{eqnarray}
X&=&X(f, \alpha)=X_{f}\ X_{\alpha},\nonumber\\
Y&=&Y(f, \alpha)=Y_{f}\ Y_{\alpha},
\end{eqnarray} 
where
\begin{eqnarray}
X_{f}&\equiv& X(f),\ \ \ 
X_{\alpha}\equiv X(\alpha),\nonumber\\
Y_{f}&\equiv& Y(f), \ \ \ 
Y_{\alpha}\equiv Y(\alpha).
\end{eqnarray}
This choice of ansatz yields the first of Eqs.(\ref{eq:constraintsvortice1}) as follows
\begin{eqnarray}
\label{eq:cons1}
\frac{n^{2} f \left(\alpha-1\right)^{2}}{r^{2}}+\frac{1}{2e^{2}}\frac{\partial V}{\partial f}&=&X_{\alpha}\ \frac{dX_{f}}{df}\ \frac{f'}{r}+X_{f}\ \frac{\alpha'}{r}\ \frac{dX_{a}}{d\alpha},\nonumber\\
&=&X_{\alpha}\ \frac{dX_{f}}{df}\ \frac{X_{f}\ X_{\alpha}}{r^{2}}+X_{f}\ \frac{dX_{a}}{d\alpha}\ Y_{f}\ Y_{\alpha},
\end{eqnarray}
where in the last line we use (\ref{eq:bogolvortices}). Identifying the terms proportional to $r^{-2}$ we obtain
\begin{equation}
\frac{1}{2}\frac{dX_{f}^{2}}{df}X_{\alpha}^{2}=n^{2} f \left(\alpha-1\right)^{2},
\end{equation}
which yields
\begin{eqnarray}
X_{\alpha}&=&\pm c_{1}\left(\alpha-1\right),\nonumber\\
X_{f}&=&\pm\frac{nf}{c_{1}},
\end{eqnarray}
for some constant $c_{1}$.

Similarly, the second of Eqs.(\ref{eq:constraintsvortice1}) can be written as
\begin{equation}
\label{eq:cons2}
\frac{2f^{2}\left(\alpha-1\right)}{r}=\frac{dY_{f}}{df}\ \frac{Y_{\alpha}\ X_{f}\ X_{\alpha}}{r}+\frac{dY_{\alpha}}{d\alpha}\ Y_{f}^{2}\ Y_{\alpha}\ r.
\end{equation}
Eq.(\ref{eq:cons2}) shows that the coefficient of the term linear in $r$ must be zero. Since $Y$ cannot be zero, the only solution is 
\begin{equation}
\label{eq:Ya}
Y_{\alpha}=c_{2},
\end{equation}
with $c_{2}$ a constant of integration. With these in hand, we can integrate $\frac{dY_{f}}{df}$ to get
\begin{equation}
Y_{f}=\mp\left(\frac{f^{2}}{c_{2}n}+c_{3}\right).
\end{equation}
Finally, the Higgs potential can be determined from the leftover condition in (\ref{eq:cons1}),
\begin{equation}
\label{eq:dV}
\frac{1}{2e^{2}}\frac{\partial V}{\partial f}=c_{3}\ n\ f\left(\frac{f^{2}}{c_{3}n}+c_{4}\right).
\end{equation}
Before integrating it, let us determine $c_{4}$ by imposing the appropriate boundary condition. Since we expect the potential to represent spontaneous symmetry breaking, its vacua at $f=1$ should be extremum, {\it i.e.}, $\frac{\partial V}{\partial f}\bigg |_{f=1}=0$. This gives $c_{4}=\frac{-1}{c_{3}n}$, or
\begin{equation}
\frac{\partial V}{\partial f}=2e^{2}f\left(f^{2}-1\right).
\end{equation}
Integrating it, we obtain
\begin{equation}
\label{eq:potvortices}
V=\frac{e^{2}}{2}\left(f^{2}-1\right)^{2},
\end{equation}
where the constant of integration is set to be zero. Potential (\ref{eq:potvortices}) is the Higgs potential that gives rise to BPS vortex defects. Notice that the Bogomol'nyi condition, $\beta\equiv 2e^{2}/\lambda=1$, is automatically satisfied. The Bogomol'nyi equations, (\ref{eq:bogolvortices}), become
\begin{eqnarray}
\label{eq:vorticesfinal}
 rf'&=&\pm nf\left(1-\alpha\right),\nonumber\\
\frac{\alpha'}{r}&=&\mp\frac{1}{n}\left(f^{2}-1\right).
\end{eqnarray}
 
The boundary value of static energy is saturated by the BPS equations. Using (\ref{eq:vorticesfinal}) we can write Eq.(\ref{eq:Evortex}) as follows
\begin{eqnarray}
 E&=&2\pi\int dr\left[\pm 2n~f(1-\alpha) f'\mp n(f^2-1)\alpha'\right].
\end{eqnarray}
we can define a functional
\begin{equation}
Q\equiv\pm 2\pi n(f^2-1)(1-\alpha), 
\end{equation}
such that the bounded value of the static energy is given by
\begin{eqnarray}
\label{eq:BPSEnergyVortex}
E&=2\pi&\int^{r=\infty}_{r=0} dQ=\pm 2\pi n\left[(f(\infty)^2-1)(1-\alpha(\infty))-(f(0)^2-1)(1-\alpha(0))\right]\nonumber\\
&=&\pm 2\pi n,
\end{eqnarray}
where $n\geq 0$ for positive sign and $n<0$ for negative sign in order to have non-negative energy.

 \subsection{Monopoles}

Monopoles are three-codimensional point defects that arise when the corresponding vacuum manifold is non-contractible. The simplest theory that has magnetic monopole in its spectrum is the Yang-Mills-Higgs theory with triplet scalar field, whose Lagrangian (for example, in~\cite{weinberg}) is
\begin{equation}
{\cal L}=\frac{1}{2}\left(D_{\mu}\phi^{a}\right)\left(D^{\mu}\phi^{a}\right)-\frac{1}{4}F_{\mu\nu}^{a}F^{a\mu\nu}-V(\phi^a\phi^a).
\end{equation}
For unit magnetic charge, the most general spherically symmetric ansatz is the so-called {\it hegdehog}:
\begin{eqnarray}
\phi^{a}&=&\hat{r}^{a}h(r),\nonumber\\
A^{a}_{i}&=&\epsilon^{aim}\hat{r}^{m}\left(\frac{1-u(r)}{er}\right),\ \ \ A^{a}_{0}=0.
\end{eqnarray}
The static energy (after rescaling) and the field equations are, respectively,
\begin{equation}
\label{eq:Emonopole}
E=\frac{4\pi\eta}{e}\int dr\ r^{2}\left(\frac{1}{2}h'^{2}+\frac{u'^{2}}{r^{2}}+\frac{\left(u^{2}-1\right)^{2}}{2r^{4}}+\frac{u^{2}h^{2}}{r^{2}}+\frac{1}{e^{2}}V(h)\right),
\end{equation}
\begin{eqnarray}
\frac{\left(r^{2}h'\right)'}{r^{2}}&=&\frac{2u^{2}h}{r^{2}}+\frac{1}{e^{2}}\frac{\partial V}{\partial h},\nonumber\\
\left(u'\right)'&=&\frac{u\left(u^{2}-1\right)}{r^{2}}+uh^{2},
\end{eqnarray} 
where, as before, we keep the potential unspecified. Applying the same prescription, we re-write them as
\begin{eqnarray}
\frac{\left(r^{2}h'-X\right)'}{r^{2}}+\frac{X'}{r^{2}}&=&\frac{2u^{2}h}{r^{2}}-\frac{1}{2}\frac{\partial V}{\partial h},\nonumber\\
\left(u'-Y\right)'+Y'&=&\frac{u\left(u^{2}-1\right)}{r^{2}}-uh^{2},
\end{eqnarray}
for $X=X(h,u)$ and $Y=Y(h,u)$. We can identify the Bogomol'nyi equations as follows
\begin{eqnarray}
\label{eq:bogolmonop}
r^{2}h'&=&X,\nonumber\\
u'&=&Y,
\end{eqnarray}
supplemented by constraint equations
\begin{eqnarray}
\frac{X'}{r^{2}}&=&\frac{2u^{2}h}{r^{2}}+\frac{1}{e^{2}}\frac{\partial V}{\partial h},\nonumber\\
Y'&=&\frac{u\left(u^{2}-1\right)}{r^{2}}+uh^{2}.
\end{eqnarray}

It can be shown, by choosing ansatz for $X$ and $Y$ which satisfy separation of variables and making use of (\ref{eq:bogolmonop}), that the first constraint condition yields
\begin{equation}
\frac{1}{2} \frac{dX_{h}^{2}}{dh}\ \frac{X_{u}^{2}}{r^{4}}+\frac{dX_{u}}{du}\frac{Y_{u}Y_{h}X_{h}}{r^{2}}=\frac{2u^{2}h}{r^{2}}+\frac{1}{e^{2}}\frac{\partial V}{\partial h}.
\end{equation}
Obviously, $\frac{\partial V}{\partial h}=0$. Therefore, without loss of generality, we can set 
\begin{equation}
V=0,
\end{equation}
the Bogomol'nyi condition for BPS solutions to exist. Also, we have $\frac{dX_{h}^{2}}{dh}X_{u}^{2}=0$. The only acceptable solution is
\begin{equation}
X_{h}=const.\equiv\pm c_{1}.
\end{equation}
The rest of the term reads
\begin{equation}
c_{1}\frac{dX_{u}}{du}Y_{u}Y_{h}=2u^{2}h,
\end{equation}
which, by means of separation of variable, dictates that
\begin{equation}
\label{eq:yhmonop}
Y_{h}=c_{2}h,
\end{equation}
and 
\begin{equation}
\label{eq:yudXu}
c_{1}Y_{u}\frac{dX_{u}}{du}=\frac{2u^{2}}{c_{2}},
\end{equation}
for some constant $c_{2}$.

The second constraint gives
\begin{equation}
\label{eq:2ndconsmonop}
\frac{dY_{h}}{dh}\ \frac{X_{h}Y_{u}X_{u}}{r^{2}}+\frac{1}{2}Y_{h}^{2}\ \frac{dY_{u}^{2}}{du}=\frac{u\left(u^{2}-1\right)}{r^{2}}+uh^{2}.
\end{equation}
It is easy to see that we can identify $Y_{h}^{2}Y_{u}\frac{dY_{u}}{du}=uh^{2}$, which, upon substitution from (\ref{eq:yhmonop}), yields
\begin{equation}
Y_{u}=\pm\sqrt{\frac{u^{2}}{c_{2}^{2}}+c_{3}},
\end{equation}
Equipped with this result, we can identify the coeficients proportional to $r^{-2}$ in (\ref{eq:2ndconsmonop}) and extract
\begin{equation}
X_{u}=\pm\frac{u\left(u^{2}-1\right)}{c_{1}\sqrt{u^{2}+c_{2}c_{3}}}.
\end{equation}
Finally, we can determine $c_{23}\equiv c_{2}c_{3}$ using (\ref{eq:yudXu}), which yields
\begin{equation}
3u^{2}-1-\frac{u^{2}\left(u^{2}-1\right)}{\left(u^{2}+c_{23}\right)}=2u^{2}.
\end{equation}
The readers may convince themselves that the above equation can consistently be satisfied provided we choose $c_{23}=0$. Hence, Eqs.(\ref{eq:bogolmonop}) are
\begin{eqnarray}
\label{eq:BPSmonop}
r^{2}h'&=&\pm\left(u^{2}-1\right),\nonumber\\
u'&=&\pm u h.
\end{eqnarray}

Inserting the BPS equations, (\ref{eq:BPSmonop}), into the static energy (\ref{eq:Emonopole}) leads to
\begin{eqnarray}
E&=&\frac{4\pi\eta}{e}\int dr\ \left(\pm\left(u^{2}-1\right)h'\pm 2uh\ u'\right)\nonumber\\
&=&\pm\frac{4\pi\eta}{e}\left(u^{2}-1\right)h\ \bigg |^{\infty}_{0}\nonumber\\
&=&\frac{4\pi\eta}{e},
\end{eqnarray}
where, in order to obtain the last line, we pick the lower sign. Thus the BPS equations (\ref{eq:BPSmonop}) with charge $n=+1$ is
\begin{eqnarray}
r^{2}h'&=&\left(1-u^{2}\right),\nonumber\\
u'&=&-u h.
\end{eqnarray}

\section{The Case for Dirac-Born-Infeld (DBI) Defects}

Finding BPS equations to the canonical defects using our method is laborious and unilluminating, since we know that they can be obtained in much shorter steps by the Bogomolnyi trick (the off-shell method). The real test to the on-shell method will be to the noncanonical defects. This section is devoted to this investigation. In particular, we will focus only on DBI domain walls and vortices. The computations for other noncanonical BPS equations will appear in the forthcoming publications. 

 \subsection{DBI Domain Wall}

Perhaps the simplest of DBI defects is the DBI domain wall, whose Lagrangian is~\cite{yisong, rubiera} 
\begin{equation}
{\cal L}=b^{2}\left(1-\sqrt{1-\frac{1}{b^{2}}\partial_{\mu}\phi\partial^{\mu}\phi}\right)-V(\phi),
\end{equation}
where $b$ is the Born-Infeld coupling constant. Ordinary domain walls can be recovered should we take $b\gg 1$. \handhika{The Hamiltonian density is
\begin{equation}
\label{eq:EDBIkink}
{\cal H}=\left[b^{2}\left(\sqrt{1+\frac{1}{b^{2}}\phi'^{2}}-1\right)+V(\phi)\right]
\end{equation}
}
from which the equation of motion follows
\begin{equation}
\left(\frac{\phi'}{\sqrt{1+\frac{1}{b^{2}}\phi'^{2}}}\right)'=\frac{\partial V}{\partial\phi}.
\end{equation}

According to our prescription, we can re-write
\begin{equation}
\left(\frac{\phi'}{\sqrt{1+\frac{1}{b^{2}}\phi'^{2}}}-X\right)'+X'=\frac{\partial V}{\partial\phi}.
\end{equation}
The Bogomol'nyi equation, therefore, is
\begin{equation}
\label{eq:bogoldbidomain}
\phi'=\frac{X}{\sqrt{1-\frac{X^{2}}{b^{2}}}}.
\end{equation}
To determine the constraint function $X$, we need to solve 
\begin{equation}
\frac{X}{\sqrt{1-\frac{X^{2}}{b^{2}}}} \frac{dX}{d\phi}=\frac{dV}{d\phi}.
\end{equation}
Integrating it, we obtain
\begin{equation}
\frac{-V}{b^{2}}+c=\sqrt{1-\frac{X^{2}}{b^{2}}}.
\end{equation}
Imposing $V=0$ at $X=0$~\cite{yisong}, we obtain
\begin{equation}
V(\phi)=b^{2}\left(1-\sqrt{1-\frac{X^{2}}{b^{2}}}\right),
\end{equation}
and Eq.(\ref{eq:bogoldbidomain}) becomes
\begin{equation}
\phi'=\pm\frac{b\ \sqrt{1-\left(1-\frac{V}{b^{2}}\right)^{2}}}{1-\frac{V}{b^{2}}}.
\end{equation}
We can see that, as in the case of canonical domain walls, BPS DBI domain walls can exist for any arbitrary potential, a feature not found in any higher-codimension defects. 

Inserting the BPS results into (\ref{eq:EDBIkink}) yields
\begin{eqnarray}
{\cal H}&=&\frac{\phi'^{2}}{\sqrt{1+\frac{1}{b^{2}}\phi'^{2}}}=b^{2}\phi'\sqrt{P(\phi)},
\end{eqnarray} 
where $P(\phi)\equiv1-\left(1-\frac{V}{b^{2}}\right)^{2}$. Thus the topological charge can be defined as
\begin{equation}
Q\equiv\int d\phi\sqrt{1-\left(1-\frac{V}{b^{2}}\right)^{2}}.
\end{equation}

 \subsection{DBI Vortex}
 
The Bogomol'nyi equations for DBI vortex was first discussed in~\cite{Shiraishi}, where they use the following field equations
\begin{eqnarray}
\label{eq:fieldDBIvortex}
\frac{\left(rf'\right)'}{r}&=&\frac{n\ f\ \alpha^{2}}{r^{2}}+\frac{1}{2}\frac{\partial V}{\partial f},\nonumber\\
\left[\left(1+\left(\frac{n\ \alpha'}{b\ r}\right)^{2}\right)^{-1/2}\frac{\alpha'}{r}\right]'&=&\frac{2\ f^{2}\ \alpha}{r},
\end{eqnarray} 
for $f$ and $\alpha$ the Higgs and gauge fields, respectively. They are obtained by varying the static energy
\begin{equation}
\label{eq:EDBIvortex}
E=2\pi\eta^{2}\int dr\ r\ \left[b^{2}\left(\sqrt{1+\left(\frac{n\alpha'}{br}\right)^{2}}-1\right)+f'^{2}+\frac{n^{2}\alpha'^{2}f^{2}}{r^{2}}+V(f)\right].
\end{equation}

The first-order equations are
\begin{eqnarray}
\label{eq:bogoldbivortex}
rf'&=&X,\nonumber\\
\frac{\alpha'}{r}&=&\frac{Y}{\sqrt{1-\frac{n^{2}}{b^{2}}Y^{2}}},
\end{eqnarray}
with
\begin{eqnarray}
\label{eq:consdbivortex}
\frac{X'}{r}&=&\frac{n\ f\ \alpha^{2}}{r^{2}}+\frac{1}{2}\frac{\partial V}{\partial f},\nonumber\\
Y'&=&\frac{2\ f^{2}\ \alpha}{r}.
\end{eqnarray}
 
 Eqs.(\ref{eq:consdbivortex}), by means of (\ref{eq:bogoldbivortex}), lead to
 \begin{equation}
 \label{eq:lable1}
 \frac{dX_{f}}{df}\ \frac{X_{f}\ X_{\alpha}^{2}}{r^{2}}+\frac{dX_{\alpha}}{d\alpha}\ \frac{X_{f}\ Y_{f}\ Y_{\alpha}}{\sqrt{1-\frac{n^{2}}{b^{2}}Y_{\alpha}^{2}\ Y_{f}^{2}}}=\frac{n^{2}\ f\ \alpha^{2}}{r^{2}}+\frac{1}{2}\frac{dV}{df},
 \end{equation}
 and
 \begin{equation}
 \label{eq:lable2}
 \frac{dY_{f}}{df}\ Y_{\alpha}\ f'+\frac{dY_{\alpha}}{d\alpha}\ Y_{f}\ \alpha'=\frac{2\ f^{2}\ \alpha}{r}. 
 \end{equation}

We can identify, from  (\ref{eq:lable1}),
\begin{eqnarray}
\label{eq:lable3}
\frac{dX_{f}}{df}\ X_{f}^{2}\ X_{\alpha}^{2}&=&n^{2}f\ \alpha^{2},\nonumber\\
\frac{dX_{\alpha}}{d\alpha}\ \frac{X_{f}\ Y_{f}\ Y_{\alpha}}{\sqrt{1-\frac{n^{2}}{b^{2}}Y_{f}^{2}\ Y_{\alpha}^{2}}}&=&\frac{1}{2}\frac{dV}{df}.
\end{eqnarray}
While from (\ref{eq:lable2}) we obtain 
\begin{eqnarray}
\label{eq:lable4}
\frac{dY_{f}}{df}\ Y_{\alpha}\ X_{f}\ X_{\alpha}&=&2\ f^{2}\ \alpha,\nonumber\\
\frac{dY_{\alpha}}{d\alpha}\ \frac{Y_{f}^{2}\ Y_{\alpha}^{2}}{\sqrt{1-\frac{n^{2}}{b^{2}}Y_{f}^{2}\ Y_{\alpha}^{2}}}&=&0.
\end{eqnarray}
The last of (\ref{eq:lable4}) yields
\begin{equation}
\label{eq:Yadbi}
Y_{\alpha}= c_{1},
\end{equation}
while from (\ref{eq:lable3}) we can collect
\begin{eqnarray}
\label{eq:Xdbi}
X_{\alpha}&=& c_{2}\ \alpha,\nonumber\\
X_{f}&=&\sqrt{n^{2}\ f^{2}+c_{3}},
\end{eqnarray}
where $c_{1}$, $c_{2}$ and $c_{3}$ are constants of integration.

Inserting (\ref{eq:Yadbi}) and (\ref{eq:Xdbi}) into the first of (\ref{eq:lable4}) gives us
\begin{eqnarray}
Y_{f}&=&\frac{2}{c_{1}\ c_{2}}\int \frac{f^{2}}{\sqrt{n^{2}\ f^{2}+c_{3}}}\ df\nonumber\\
&=&\frac{1}{c_{1}c_{2} n^{2}} f \sqrt{n^{2} f^{2}+c_{3}}-\frac{c_{3}}{c_{1} c_{2} n^{3}}\ln\left(2 n^{2} f+2 n \sqrt{n^{2} f^{2}+c_{3}}\right).
\end{eqnarray}
As in the case of ordinary vortices above, the problem can be simplified should we take $c_{3}=0$. We now have
\begin{eqnarray}
X_{f}&=&n\ f,\nonumber\\
Y_{f}&=&\frac{2}{c_{1}\ c_{2}\ n}\int f\ df=\frac{f^{2}}{c_{1}\ c_{2}\ n}+c_{4}.
\end{eqnarray}

The second of (\ref{eq:lable3}) now reads
\begin{equation}
c_{1}c_{2}\frac{nf\left(\frac{f^{2}}{c_{1}c_{2}n}+c_{4}\right)}{\sqrt{1-\frac{c_{1}^{2}n^{2}}{b^{2}}\left(\frac{f^{2}}{c_{1}c_{2}n}+c_{4}\right)^{2}}}=\frac{1}{2}\frac{dV}{df}.
\end{equation}
At $f=1$ (the vacua of the potential) we require~\cite{Shiraishi}
\begin{equation}
\frac{dV}{df}\bigg |_{f=1}=0,
\end{equation}
which imposes $c_{4}=-\frac{1}{c_{1}c_{2}n}$. We then have
\begin{equation}
\frac{dV}{df}=\frac{2\ f\left(f^{2}-1\right)}{\sqrt{1-\frac{1}{c_{2}^{2}b^{2}}\left(f^{2}-1\right)^{2}}},
\end{equation}
which, upon integration and setting the constant of integration such that the vanishing potential is located at $f=1$, results in
\begin{equation}
\label{eq:dbivortexpoten}
V(f)=b^{2}\left(1-\sqrt{1-\frac{1}{b^{2}}\left(f^{2}-1\right)^{2}}\right),
\end{equation}
where we absorb $c_{2}$ into $b$.

Eqs.(\ref{eq:bogoldbivortex}) become
\begin{eqnarray}
\label{eq:bogoldbivortexfinal}
rf'&=&nf\alpha,\nonumber\\
\frac{\alpha'}{r}&=&\frac{f^{2}-1}{n\sqrt{1-\frac{1}{b^{2}}\left(f^{2}-1\right)^{2}}},
\end{eqnarray}
by once again absorbing $c_{2}$ into $n$ now. Eqs.(\ref{eq:bogoldbivortexfinal}) constitute the corresponding Bogomol'nyi equations for DBI vortices which satisfy (\ref{eq:fieldDBIvortex}) provided the Higgs potential takes the form of Eq.(\ref{eq:dbivortexpoten}).

As in the previous cases, the BPS conditions directly saturates the energy bound (\ref{eq:EDBIvortex}),
\begin{equation}
E=2\pi\eta^{2}\int dr\ \left(2Xf'+\frac{n^{2}Y}{\sqrt{1+\left(\frac{nY}{b}\right)^{2}}}\alpha'\right)=2\pi\eta^{2}\int dQ,
\end{equation}
whose charge is given by
\begin{eqnarray}
Q&\equiv&\left[n\left(f^{2}-1\right)\alpha\right]\bigg |^{f(\infty),\alpha(\infty)}_{f(0),\alpha(0)}\nonumber\\
&=&n.
\end{eqnarray}
Note that the minimum energy is the same as in the ordinary vortex case, Eq.(\ref{eq:BPSEnergyVortex}). This is because the DBI-modification does not affect the vacuum topology, since the DBI form changes only the UV-regime behavior of the solutions while the topology is the near-infinity (infra-red) property of the field.

\section{Conclusions} 

Bogomol'nyi equations usually can be obtained by completing the square of the terms in the corresponding static energy. In most cases it is difficult to do, due to the complexity of the kinetic term(s) of the Lagrangian. Here, we offer an alternative way of constructing the first-order differential equations by manipulating the full second-order equations based on what we call the on-shell method. We showed that the known Bogomol'nyi equations for domain walls, vortices, and monopoles can be derived by choosing certain auxiliary functions, constants of integration, and the potential coupling constants. Later we applied this method to the DBI defects, and reproduced the known results. Our results confirm the well-known results that, except for the domain walls, the canonical and DBI defects can have BPS solutions only for some unique forms of the potential. 

Our method does not start from the Hamiltonian but instead considering the Euler-Lagrange equations directly. We can compare it with the {\it first-order framework} for generalized defects discussed in~\cite{BazeiaLosano1, BazeiaLosano2}. There, the authors derive the first-order BPS equations from the Euler-Lagrange equations by imposing the {\it stressless condition}; {\it i.e.,} that the stress components of the energy-momentum tensor vanish. This construction is interesting in its own right, since it reveals the relation between BPS equations and the vanishing condition of the stress tensor. Yet, despite the apparent similarity between their method and ours, we believe our method is genuine in the sense that we do not take into account the energy consideration whatsoever. Also, ours is more general since it works on any defects in higher codimensions, while the stressless-condition method so far has been applied only to codimensional-one. In any case, it is interesting to investigate whether these two approaches could be combined to produce a better construction for obtaining the first-order BPS equations.

In general, our conjecture for solving Eq.(\ref{eq:algebraic}) could produce more algebraic equations than the number of auxiliary functions and this may result in the inconsistency of the solutions. It is somehow very surprising that our conjecture reproduces the well-known Bogomol'ny equations with no inconsistency. We are not trying to argue that our on-shell method will give all first-order differential (Bogomol'nyi) equations which are subset of the corresponding second-order differential (Euler-Lagrange) equations as we have no prove of it, neither do we provide a method to solve the Bogomol'nyi equations. Moreover, this on-shell method opens up the possibility for non-separation-of-variable ansatz. If exists, the meaning of these ``other" BPS solutions is unclear to us. We admit that our method is not fully complete and there are probably few steps in our on-shell method that may not have strong mathematical or physical grounds, especially our conjecture about the algebraic equations. However, we have not found so far any contradiction with the well-known results in the literature and we tend to believe that we are on the right track.

As stressed before, this work is still in its preliminary stage. Much still needs to be done. In particular, we have yet presented the result for the DBI monopoles case. The non-Abelian and the DBI natures of the defects~\cite{grandi} make it difficult to obtain the auxiliary constraint functions. The investigation is still going on, and the result  will appear in the forthcoming publications~\cite{dhikaardian}. Modifications to our on-shell method could also be done for example by taking $X\equiv X(r,\phi)$ or by imposing other condition to the auxiliary fields rather than taking them as separable functions.

It is known that in the supersymmetric extension there are actually more solutions (or states) to the supersymmetric transformation of the fields that break the same amount of supersymmetry but in general may not be solutions of the Euler-Lagrange equations, see~\cite{weinberg}. It might also be interesting to see relation between the Bogomol'ny solutions produced by our on-shell method with the supersymmetric solutions (or states). From cosmological point of view, our work finds its relevance the most perhaps in the study of $k$-defects~\cite{babichev1, sarangi, babichev2, pavlovsky}. It is unclear whether BPS solutions in ordinary defects still have the minimum energy in the noncanonical cases; or whether it even exists. This method might shed a new light on finding such minimum-energy states. This may have interesting cosmological implications in the study of defects in the very early universe. One may also try to apply our on-shell method to finding Bogomol'nyi equations of the theories where the background metric is not Minkowskian but is in general curved, for example as in~\cite{comtet, gibbons1, gibbons2, horvathy}. Since the on-shell method does not deal with the Hamiltonian, we believe it is more suitable for such a purpose since the notion of {\it energy} is rather subtle in the gravitational theories. Another direct application in gravity is to find solutions of Einstein equations\footnote{An interesting reverse-proposal was proposed by Singleton~\cite{singleton} who applies exact method in obtaining analytic solutions in general relativity for solving Yang-Mills equations.}. Extension of our method to more than one parameter dependent is necessary especially for stationary solution of gravity such as in Kerr black holes where the metrics depend on radial and angular coordinates. We hope to elaborate them more in the future.

\acknowledgments

We are indebted to Laksana Tri Handoko, Eduardo da Hora, Peter Horvathy, Terry Mart, Kiyoshi Siraishi, Anto Sulaksono, Andrzej Wereszczynski, and Yisong Yang for enlightening discussions. This work is partially funded by the Competitive Grant-LIPI 2013 on ``Fluida in High Energy". A.N.A acknowledges University of Malaya for the support through the University of Malaya Research Grant (UMRG) Programme RP006C-13AFR and RP012D-13AFR. H.S.R acknowledges support from University of Indonesia through Research Cluster Grant 2014 on ``Non-perturbative Phenomena in Nuclear Astrophysics and Cosmology" No.~1709/H2.R12/HKP.05.00/2014, and thanks Riemann Workshop on ``Gauge Theories in Higher Dimensions" at the Gottfried Wilhelm Leibniz Universitat Hannover, Germany, for the hospitality during the completion of this work.


\end{document}